\documentclass[10pt]{article}
\usepackage[dvips]{graphicx}

\usepackage{epsfig}
\usepackage{graphicx}
\usepackage{indentfirst}
\usepackage{amsmath}
\usepackage{amsfonts}
\usepackage{amssymb}

\begin{document}

\begin{center}
{\Large\bf  Evaporation phenomena in f(T) gravity\\}
\medskip
M. J. S. Houndjo$^{(a,b)}$\footnote{e-mail:
sthoundjo@yahoo.fr}, D. Momeni\footnote{e-mail: d.momeni@yahoo.com}$^{(c)}$,\, R. Myrzakulov$^{(c)}$\footnote{e-mail:rmyrzakulov@csufresno.edu} and M. E. Rodrigues$^{(d)}$\footnote{e-mail: esialg@gmail.com}\\
$^{a}${ \it Departamento de Engenharia e Ci\^{e}ncias Naturais- CEUNES -
Universidade Federal do Esp\'irito Santo\\
CEP 29933-415 - S\~ao Mateus - ES, Brazil}\\
 $^{b}${\it Institut de Math\'{e}matiques et de Sciences Physiques (IMSP) - 01 BP 613 Porto-Novo, B\'{e}nin}\\
$^{c}${\it
Eurasian International Center for Theoretical Physics -
 Eurasian National University,\\ Astana 010008, Kazakhstan}\\
$^{d}$\,{\it\ Faculdade de F\'{\i}sica, Universidade Federal do Par\'{a}, 66075-110, Bel\'em, Par\'{a}, Brazil}
\
\date{}

\end{center}
\begin{abstract}
We formulate evaporation phenomena in a generic model of generalized teleparallel gravity in Weitzenbock spacetime with diagonal and non-diagonal tetrads basis. We also perform the perturbation analysis around the constant torsion scalar solution named Nariai spacetime which is an exact solution of field equations as the limiting case of the Schwarzschild-de Sitter and in the limit where two back hole and their cosmological horizons coincide. By a carefully analysis of the horizon perturbation equation, we show that (anti)evaporation  can not happen if we use a diagonal tetrad basis. This result implies that a typical black hole in any generic form of generalized teleparallel gravity is frozen in its initial state if we use the diagonal tetrads. But in the case of non-diagonal tetrads the analysis is completely different. By a suitable non trivial non-diagonal tetrad basis we investigate the linear stability of the model under perturbations of the metric and torsion simultaneously. We observe that in spite of the diagonal case, both evaporation and anti evaporation can happen. The phenomena depend on the initial phase of the horizon perturbation. In the first mode when we restrict ourselves to the first lower modes the (anti)evaporation happens. So, in non-diagonal case the physical phenomena is reasonable. This is an important advantage of using non-diagonal tetrads instead of the diagonal ones. We also see that this is an universal feature, completely independent from the form of the model.
\end{abstract}

Pacs numbers: 04.50. Kd, 04.70.Dy
\section{Introduction}
Black holes as thermodynamic systems have temperature, entropy and heat capacity. All kinds of the thermodynamic laws are satisfied if we appropriately define and identify surface gravity $\kappa$, the temperature $T$, the area $A$ as well as the entropy $S$. Mass as the total energy of the system can be used to calculate the heat capacity and a full consistent description of stability conditions and also the mechanism of phase transition from stable to unstable cases in the presence of the charges \cite{davis}. Radiation of the black hole as a black body system has the same spectrum as the Planck prediction for photons and can be verified  widely with different methods. This radiation evaporates the surface of black hole with a thermal spectrum exactly the same as the Planck black body prediction. The key object here is to label surface gravity by temperature. But if this black hole is located in a de-Sitter or Anti-de Sitter background, it means that it was a black hole in cosmology, then it can be shown that both evaporation and anti evaporation can happen if the black hole is slightly perturbed from the background. The essence of evaporation or anti evaporation depends on how the radius of horizon evolves during the perturbation. In Einstein gravity it is a well known result and it has been proved by analytical method in s-wave approximation of a single mode perturbation \cite{Bousso:1997wi}  and numerically by including higher modes. 
\par
In modified gravity (see for a review \cite{Nojiri:2010wj}-\cite{Nojiri:2006ri}), it seems that for some kinds of $f(R)$ gravity models as the very popular modification of Einstein gravity, such anti evaporation happens \cite{Nojiri:2013su}. The first step is how to model the spacetime of Schwarzschild-de Sitter (SdS) in a limiting case in which the final state is a black hole with one horizon (degenerated) and in thermal equilibrium to use the equilibrium thermodynamics. The solution of Einstein equation in the limiting case of SdS is proposed by Nariai \cite{Nariai}. This solution which we propose to investigate in details in next sections, is a vacuum inhomogeneous solution of Einstein field equations in the presence of a cosmological constant. It has different representations in static and non static patches. So, the perturbation on this spacetime can be done in two patches.
\par
As an alternative to Einstein gravity and in spite of $f(R)$ gravity, a generalized theory (gauge theory) is recently proposed for gravity on Weitzenbock spacetime in which the field equations are  second order and are well defined by a suitable choice of tetrad basis \cite{fT}-\cite{Rodrigues:2013uua}. Our aim in this paper is to study the (anti)evaporation effect in $f(T)$ gravity by studying the perturbation of metric around a Nariai spacetime.  The paper is presented as follows: In section $2$ the foundations of $f(T)$ gravity has been reviewed. In section $3$, we derived the field equation for a Nariai's like non static inhomogeneous spacetime in Weitzenboch approach to gravity. In section $4$, we perform the perturbation analysis around the Nariai solution and find the horizon perturbation function. In section $5$ we apply a non-diagonal tetrads basis and  compute the first order perturbations. We summarize and conclude in the final section. 
\section{\large The generality on $f(T)$ theory}
We start by the basis of the generalized teleparallel gravity, named  $f(T)$ theory. The line element in a general spacetime is defined as, 
\begin{equation}
dS^{2}=g_{\mu\nu}dx^{\mu}dx^{\nu}\; .\label{david1}
\end{equation} 
By using the representation of the tetrad matrix, the metric can be projected in the tangent space to the manifold where we write the line element as
\begin{eqnarray}
dS^{2} &=&g_{\mu\nu}dx^{\mu}dx^{\nu}=\eta_{ij}\theta^{i}\theta^{j}\label{david2}\; ,\\
dx^{\mu}& =&e_{i}^{\;\;\mu}\theta^{i}\; , \; \theta^{i}=e^{i}_{\;\;\mu}dx^{\mu}\label{david3}\; ,
\end{eqnarray} 
with $\eta_{ij}=diag[-1,1,1,1]$ and $e_{i}^{\;\;\mu}e^{i}_{\;\;\nu}=\delta^{\mu}_{\nu}$ or  $e_{i}^{\;\;\mu}e^{j}_{\;\;\mu}=\delta^{j}_{i}$. One can determine the square root of the metric determinant as  $\sqrt{-g}=\det{\left[e^{i}_{\;\;\mu}\right]}=e$. The main description of the spacetime can be done through the tetrad matrix essentially based on the Weitzenbock's connection defined as 
\begin{eqnarray}
\Gamma^{\alpha}_{\mu\nu}=e_{i}^{\;\;\alpha}\partial_{\nu}e^{i}_{\;\;\mu}=-e^{i}_{\;\;\mu}\partial_{\nu}e_{i}^{\;\;\alpha}\label{david4}\; .
\end{eqnarray}
The torsion tensor is defined as
\begin{eqnarray}
T^{\alpha}_{\;\;\mu\nu}&=&\Gamma^{\alpha}_{\nu\mu}-\Gamma^{\alpha}_{\mu\nu}=e_{i}^{\;\;\alpha}\left(\partial_{\mu} e^{i}_{\;\;\nu}-\partial_{\nu} e^{i}_{\;\;\mu}\right)\label{david5}\;,
\end{eqnarray}
and the associated contorsion as
\begin{eqnarray}
K^{\mu\nu}_{\;\;\;\;\alpha}&=&-\frac{1}{2}\left(T^{\mu\nu}_{\;\;\;\;\alpha}-T^{\nu\mu}_{\;\;\;\;\alpha}-T_{\alpha}^{\;\;\mu\nu}\right)\label{david6}\,.
\end{eqnarray}
We also define a tensor $S_{\alpha}^{\;\;\mu\nu}$ from the torsion and contorsion as
\begin{eqnarray}
S_{\alpha}^{\;\;\mu\nu}&=&\frac{1}{2}\left( K_{\;\;\;\;\alpha}^{\mu\nu}+\delta^{\mu}_{\alpha}T^{\beta\nu}_{\;\;\;\;\beta}-\delta^{\nu}_{\alpha}T^{\beta\mu}_{\;\;\;\;\beta}\right)\label{david7}\;.
\end{eqnarray}
By using (\ref{david5})-(\ref{david7}), we define the scalar torsion  as  
\begin{eqnarray}
T=T^{\alpha}_{\;\;\mu\nu}S^{\;\;\mu\nu}_{\alpha}\label{david8}\; .
\end{eqnarray}
Let us write the action as  \cite{fT},
\begin{eqnarray}
S[e^{i}_{\mu},\Phi]=\int\; d^{4}x\;e\left[\frac{1}{16\pi}f(T)+\mathcal{L}_{Matter}\left(\Phi\right)\right]\label{david9}\; ,
\end{eqnarray}
where the units $G=c=1$ are used and  $\Phi$ denotes matter fields. By assuming the action (\ref{david9}) as a functional of the tetrad field  $e^{i}_{\mu}$ and matter field $\Phi$,   and vanishing its variation with respect to $e^{i}_{\nu}$, one gets  the following field equation \cite{fT},
\begin{eqnarray}
S^{\;\;\nu\rho}_{\mu}\partial_{\rho}Tf_{TT}+\left[e^{-1}e^{i}_{\mu}\partial_{\rho}\left(ee^{\;\;\alpha}_{i}S^{\;\;\nu\rho}_{\alpha}\right)+T^{\alpha}_{\;\;\lambda\mu}S^{\;\;\nu\lambda}_{\alpha}\right]f_{T}+\frac{1}{4}\delta^{\nu}_{\mu}f=4\pi\mathcal{T}^{\nu}_{\mu}\label{david10}\; ,
\end{eqnarray}
where $\mathcal{T}^{\nu}_{\mu}$ denotes the energy momentum tensor, with $f_{T}=d f(T)/d T$ and $f_{TT}=d^{2} f(T)/dT^{2}$. Note that the spacial choice $f(T)=a_{1}T+a_{0}$ leads to the teleparallel equivalence of general relativity (TEGR) with a cosmological constant. In this paper we leave this trivial case and propose to search and work with the case where $f_{TT}\neq0$.
\section{On the Nariai spacetime}
In this section we investigate the dynamic of the whole universe  by assuming  that a Nariai spacetime exists for a non trivial TEGR case. One form of Nariai spacetime is described by the following  ansatz of metric\cite{Nojiri:2013su},
\begin{eqnarray}
dS^2=e^{2A(t,y)}\left(-dt^2+dy^2\right)+e^{-2B(t,y)}d\Omega^2\,,\;\;\;\; d\Omega^2\equiv d\theta^2+\sin^2\theta d\phi^2\;.
\end{eqnarray}
Here $d\Omega^2$ is the metric of a unit sphere. 
The corresponding tetrad reads
\begin{eqnarray}
 \{e^{a}_{\mu}\}=\{e^A,\,e^A,\,e^{-B},\,e^{-B}sin\theta\}.
\end{eqnarray}

According to this metric, the non-null components of the torsion tensor read
\begin{eqnarray}
T^{0}_{\;\;01}=-A'\;, \;\;T^{1}_{\;\;01}=\dot{A}\;,\;\;T^{2}_{\;\;02}=-\dot{B}\;,\;\; T^{2}_{\;\;21}=B'\;,\;\; T^{3}_{\;\;03}=-\dot{B}\;,\;\; T^{3}_{\;\;13}=-B'\;,\;\;T^{3}_{\;\;23}=\cot{\theta}\,\,,
\end{eqnarray}
while whose of the contorsion are
\begin{eqnarray}
K_{\;\;\;\;0}^{01}=-A'e^{-2A},\;K_{\;\;\;\;1}^{01}=-\dot{A}e^{-2A},\;K_{\;\;\;\;2}^{02}=K_{\;\;\;\;3}^{03}=\dot{B}e^{-2A},\nonumber\\ K_{\;\;\;\;2}^{12}=K_{\;\;\;\;3}^{13}=-B'e^{-2A},\;K_{\;\;\;\;3}^{32}=-e^{2B}\cot\theta\,\,,
\end{eqnarray}
and the non-null components of the tensor $S_{\alpha}^{\;\;\mu\nu}$ are
\begin{eqnarray}
S_{0}^{\;\;01}=-B'e^{-2A}\,,\;S_{0}^{\;\;02}=S_{1}^{\;\;12}=\frac{e^{2B}}{2}\cot{\theta}\,,\;S_{1}^{\;\;01}=-\dot{B}e^{-2A}\,,\nonumber\\  S_{2}^{\;\;02}=S_{3}^{\;\;03}=\frac{e^{-2A}}{2}\left(\dot{A}-\dot{B}\right)\,,\; S_{2}^{\;\;12}=S_{3}^{\;\;13}=\frac{e^{-2A}}{2}\left(B'-A'\right)\,\,.
\end{eqnarray}
The dot denotes the derivative with respect to time while the prime denotes the derivative with respect to the spacial coordinate $y$.\par
Making use of the above non-components, the torsion scalar can be calculated getting
\begin{eqnarray}
T=-2e^{-2A}\Big[B'^2-\dot{B}^2+2\left(\dot{A}\dot{B}-A'B'\right)\Big]\,\,.
\end{eqnarray}
We can now write down the vacuum field equations as follows
\begin{eqnarray}\label{EOMS}
e^{-2A}B'T'f_{TT}+e^{-2A}\left[B''-2B'^2+\dot{B}^2-2\dot{A}\dot{B}+\frac{1}{2}e^{2(A+B)}\right]f_{T}-\frac{f}{4}=0\,,\\
e^{-2A}\dot{B}\dot{T}f_{TT}+e^{-2A}\left[B'^2-2A'B'+\ddot{B}-2\dot{B}^2-\frac{1}{2}e^{2(A+B)}\right]f_T+\frac{f}{4}=0\,,\\
e^{-2A}\left[(A'-B')T'-(\dot{A}-\dot{B})\dot{T}\right]f_{TT}+\nonumber\\e^{-2A}\left[\ddot{B}-B''+A''-\ddot{A}+2(B'^2-\dot{B}^2+\dot{A}\dot{B}-A'B')\right]f_T+\frac{f}{2}=0\label{david19}\,,\\
B'\dot{T}f_{TT}-\left[(\dot{A}+\dot{B})B'-\dot{B}'+A'\dot{B}\right]f_T=0\, ,\label{david20}\\ 
\dot{B}T'f_{TT}-\left[(\dot{A}+\dot{B})B'-\dot{B}'+A'\dot{B}\right]f_T=0\, ,\label{david21}\\ 
\dot{T}f_{TT}\cot{\theta}=0\,,\label{david22}\\
T'f_{TT}\cot{\theta}=0\,\,.
\end{eqnarray}
Here, we see that constraints are imposed to both the torsion scalar and the algebraic function $f$ through its derivatives. However, one can carefully analyse the field equations in order to get solutions different from the TEGR one. In this paper we will work with constant torsion solutions in which $T=T_0$, for which the last two equations are satisfied identically for a generic form of $f(T)$. Also, we assume that $f_{TT}\neq0$. Another option is to work with a non-diagonal tetrads as we will see later.

\section{Nariai solution in $f(T)$-gravity}
Nariai spacetime is proposed firstly as a static, inhomogeneous spherically symmetric with cosmological constant solution of Einstein equation \cite{Nariai}. It is considered as the extremal limit of the SdS space time where the cosmological horizon and the black hole horizon coincides. Here we make a little review on the basic properties of the SdS solution. The generic form of a static spherically symmetric solution in diagonal form reads 
\begin{equation}
ds^2=-e^{\nu(r)}dt^2+e^{\lambda(r)}dr^2+r^2d\Omega^2\,,
\end{equation}
where  $e^{\nu(r)}$ for SdS is,
\begin{equation}
e^{\nu(r)}=e^{-\lambda(r)}=1-\frac{2 M}{r}-\frac{\Lambda}{3}r^2\,.\label{solution}
\end{equation}
Here, $M$ is an integration constant which can be interpreted as the mass of spacetime, and $\Lambda$ is the famous cosmological constant. This solution represents a constant curvature solution.  As we know, different $f(T)$ theories models posses such kind of solution \cite{Rodrigues:2012wt}. In $f(T)$ gravity the solution satisfies the analogue condition as $T=T_0$, where if we put it in the field equations we obtain the following constraint on the form of $f(T)$,
\begin{equation}
f'(T_0)=\frac{f(T_0)}{2\Lambda}\,,\label{dscondition}
\end{equation}
which is a consequence of Eq. (\ref{david10}) in the case of Weitzenbock's space.
We return to SdS solution given by (\ref{solution}). If we solve the cubic equation $e^{\nu(r)}=0$, for $0<9M^2\Lambda<1$, we obtain two positive roots $r_h$ and $r_{c}$, which matches two black hole and cosmological horizons. Using these parameters, the metric function of SdS can be written in the following form,
\begin{eqnarray}
e^{\nu(r)}=-\frac{\Lambda}{3r}(r-r_h)(r-r_c)(r+r_h+r_c).
\end{eqnarray}

In the case of massless black hole when $M\rightarrow0$, we again obtain the static de-Sitter spacetime. The extremal case happens when  the black hole horizon becomes more close to the cosmological horizon . In this case, the radial coordinate $r$ is  degenerated and is useless for us. To have a physically acceptable radial coordinate  which works only in the extremal limit we must introduce another radial coordinates. After an appropriate coordinate transformation 
by  going to the extremal limit, we introduce a dimensionless small parameter $\eta$ such that $9M^2\Lambda=1-3\eta^2$ where $ \eta\in(0,1]$. The Nariai cosmological metric is characterized by $\eta\rightarrow 0$. The metric is SdS spacetime in  the maximal case. The horizon's topology is  $S^1\times S^2$. This is a very interesting point that, this kind of topology is the same as the topology of a black ring, as the vacuum stationary exact solution of Einstein gravity in five dimension\cite{Emparan}. It describes the geometry of an embedded metric for two co-center spheres with the same radius. In this representation of Nariai solution, the location of singularity  is hidden under  this kind of transformation and another important point is that the metric is now free of singularity everywhere.\par

In this paper we will deal with the cosmological patch of Nariai metric. In these global coordinates  the Nariai spacetime finalizes as \cite{Lorenzo},
\begin{equation}
ds^2=-\frac{1}{\Lambda\cos^2 \tau}\left(-d\tau^2+dy^2\right)+\frac{1}{\Lambda}d\Omega^2\,,\label{Nariai}
\end{equation}
where 
\begin{equation}
\cos\tau\cosh t=1\,.\label{transf}
\end{equation}
Here time is a bounded parameter in a strip $0<\tau<\pi/2$ maps to $0<t<+\infty$. It is easy to check that (\ref{Nariai}) solves the system of equations (\ref{EOMS}) under the constraint (\ref{dscondition}). In fact, by substituting (\ref{Nariai}) in (\ref{EOMS}) we obtain the constant torsion condition (\ref{dscondition}). Therefore, as an unperturbed solution, (\ref{Nariai}) is an exact solution of $f(T)$ gravity under the constraint (\ref{dscondition}). We will  perform perturbation around (\ref{Nariai}) in the system (\ref{EOMS}).\par
Perturbations of the Nariai space-time (\ref{Nariai}) could be written in terms of $\delta A(\tau,x)$ and $\delta B(\tau,x)$,
\begin{eqnarray}
A = - \ln \left[ \sqrt{\Lambda} \cos \tau \right] + \delta A(\tau,y)\, ,\quad
B = \ln \sqrt{\Lambda} + \delta B(\tau,y)\,.
\end{eqnarray}
Then, we find
\begin{eqnarray}
\delta T =-2\Lambda\sin(2\tau)\delta\dot{B} .\label{varT}
\end{eqnarray}
Let us assume that our $f(T)$ gravity theory possesses the Nariai solution for $T=T_0$,
such that condition (\ref{dscondition}) is satisfied. At the first order,
the perturbed equations from (\ref{EOMS}) read:
\begin{eqnarray}\label{EOMS-Pert}
\delta B+\cos^2\tau\delta B''-\sin2\tau\delta\dot{B}+\frac{\delta T}{2\Lambda}(\alpha-\frac{1}{2})=0\,,\\
\frac{\delta T}{2\Lambda}(\frac{1}{2}-\alpha)+\cos^2\tau\delta\ddot{B}-\delta B=0\,,\\
-\alpha\sin\tau\cos\tau\frac{\delta\dot{T}}{2\Lambda}+\delta A+\frac{\delta T}{4\Lambda}+\sin\tau\cos\tau\delta\dot{B}+\frac{\cos^2\tau}{2}\Big[\delta\ddot{B}-\delta\ddot{ A}-(\delta B-\delta A)''\Big]=0\,,\\
f_T(\tan\tau\delta B'-\delta\dot{B}')=0.
\end{eqnarray}
Here $\alpha=\frac{\Lambda f''(T_0)}{f'(T_0)}$ and in order to avoid TEGR, we suppose $f_T\neq0$. The fourth equation of the above system can be integrated, giving
\begin{equation}
\delta B(y,t)=\frac{c_1(y)}{\cos\tau}+c_2\,.\label{deltaB}
\end{equation}
By inserting (\ref{deltaB}) in expression resulting from the sum of the two first perturbed equations, we obtain
 \begin{eqnarray}
 c_1(y)=c_0\sin(y-\theta)\label{c1}\,.
 \end{eqnarray}
 Here $\theta$ denotes the initial angle of perturbations. So, we have,
\begin{equation}
\delta B(y,t)\equiv\delta B(\tau)=c_0\sin(y-\theta)\sec\tau+c_2\label{deltaBytau}\,,
\end{equation}
where for a black hole located at $y=y_h$, the horizon is defined as 
\begin{equation}
(\nabla\delta B)^2\equiv-(\frac{\partial\delta B}{\partial \tau})^2+(\frac{\partial\delta B}{\partial y})^2=0\label{location}.
\end{equation}
Setting $\Lambda=1$, one gets
\begin{equation}
r_0(\tau)^{-2}=\mathrm{e}^{2B(\tau,y_h)}=1+\delta B(\tau)\,.
\label{relation}
\end{equation}
As we mention above, (anti)evaporation related to (decreasing) increasing values of $\delta B(\tau)$. Using (\ref{deltaBytau}), and since the location of the horizon of the black hole in $f(T)$  is
\begin{eqnarray}
y_h=\theta-\tau-\frac{\pi}{2}+n\pi,\ \ n\in\mathcal{Z}\,,
\end{eqnarray} 
 we have
\begin{eqnarray}
\delta B(\tau)\equiv\delta B(\tau,y_h) =c_0(-1)^{n+1}+c_2\,.  \label{deltaBtau}
\end{eqnarray}
 Using (\ref{deltaBtau}) in (\ref{relation}) we find 
\begin{equation}
r_0(\tau)^{-2}=1+c_0(-1)^{n+1}+c_2\,.
\end{equation}
The radius of black hole horizon in $f(T)$ (it is defined as the general relativity case by $(\nabla\delta B)^2=0$) is frozen and never diverge. It means the black hole never evaporates or anti-evaporates. It implies that the final stage of the black hole using diagonal tetrads basis in this formation process is the same as the initial radius and this result does not dependent on the form of the $f(T)$ model. 
 It indicates the absence of the evaporation in $f(T)$  scenario of gravity if we apply the formalism to the diagonal teterds. To conclude the more general statement which be independence of the form of tetrads, and because the form of the theory here is not Lorentz invariant we need to perform the analysis with non-diagonal tetrads. In this case by performing an appropriate boost transformation on the diagonal tetrad basis we can obtain a new tetrads basis in which now $f_{TT}\neq0$. It avoids to obtain the TEGR. Although our analysis in this work is based on constant torsion systems, the next section is devoted to the analysis of the evaporation phenomena in the non-diagonal tetrads basis. 

\section{Non-diagonal tetrad solutions} 

In this section we undertake a non-diagonal type of tetrad. This choice of tetrad is essentially due to the fact that the diagonal tetrad usually leads  to a constraint on the algebraic where generally the algebraic function is reduced to that of TEGR theory. In this section, we try to solve this problem by considering a type of non-diagonal tetrad.
To do so, let us consider the following non-diagonal tetrad
\begin{eqnarray}\label{nontet}
\{e^{a}_{\;\;\mu}\}=\left[\begin{array}{cccc}
e^A&0&0&0\\
0&\cos\phi\sin\theta \,e^A &\cos\phi\cos\theta \,e^{-B} &-\sin\phi\sin\theta \,e^{-B}\\
0&\sin\phi\sin\theta \,e^A &\sin\phi\cos\theta \,e^{-B}&\cos\phi\sin\theta \,e^{-B} \\
0&\cos\theta \,e^{A} &-\sin\theta \,e^{-B} &0
\end{array}\right]\;,
\end{eqnarray}  
Thereby, the torsion tensor components can be calculated giving
\begin{eqnarray}
T^{0}_{\;\;01}=-A'\;, \;\;T^{1}_{\;\;01}=\dot{A}\;,\;\;T^{2}_{\;\;02}=-\dot{B}\;,\;\; T^{2}_{\;\;21}=B'+e^{A+B}\;,\;\; T^{3}_{\;\;03}=-\dot{B}\;,\;\; T^{3}_{\;\;13}=-B'-e^{A+B}\,\,,
\end{eqnarray}
while those of the contorsion tensor reads
\begin{eqnarray}
K_{\;\;\;\;0}^{01}=A'e^{-2A},\;K_{\;\;\;\;1}^{01}=\dot{A}e^{-2A},\;K_{\;\;\;\;2}^{02}=K_{\;\;\;\;3}^{03}=-\dot{B}e^{-2A},\nonumber\\ K_{\;\;\;\;2}^{12}=K_{\;\;\;\;3}^{13}=(B'+e^{A+B})e^{-2A}\,\,,
\end{eqnarray}
and the corresponding tensor 
 $S_{\alpha}^{\;\;\mu\nu}$ are 
\begin{eqnarray}
S_{0}^{\;\;01}=-(B'+e^{A+B})e^{-2A}\,,\;S_{1}^{\;\;01}=-\dot{B}e^{-2A}\,,\nonumber\\  S_{2}^{\;\;02}=S_{3}^{\;\;03}=\frac{e^{-2A}}{2}\left(\dot{A}-\dot{B}\right)\,,\; S_{2}^{\;\;12}=S_{3}^{\;\;13}=\frac{e^{-2A}}{2}\left(B'+e^{A+B}-A'\right)\,\,.
\end{eqnarray}
 Therefore, the torsion scalar can be computed giving
 \begin{eqnarray}
 T=-2e^{2A}\left[B'^2+2\left(e^{A+B}-A'\right)B'-\dot{B}^2+2\dot{A}\dot{B}+e^{2(A+B)}
 -2A'e^{A+B}\right]\,\,\label{T2}.
 \end{eqnarray}
 Consequently, the field equations are
\begin{eqnarray}
e^{-2A}\left(B'+e^{A+B}\right)T'f_{TT}-e^{-2A}\left[2B'^2+2\dot{A}\dot{B}-(B''+\dot{B}^2)
+e^{A+B}(B'-A')\right]f_T-\frac{f}{4} =0\label{david49}\\
\left(B'+e^{A+B}\right)\dot{T}f_{TT}-\left[\dot{B}B'
-\dot{B}'+\dot{A}B'+A'\dot{B}\right]f_{T}=0\label{david50}\\
\dot{B}T'f_{TT}-\left[\dot{B}B'
-\dot{B}'+\dot{A}B'+A'\dot{B}\right]f_T=0\\
e^{-2A}\dot{B}\dot{T}f_{TT}+e^{-2A}\Big[B'^2-2A'B'+\ddot{B}-2\dot{B}^2
+e^{A+B}\left(B'-A'\right)\Big]f_{T}+\frac{f}{4}=0\\
\frac{1}{2}e^{-2A}\left[(A'-B'-e^{A+B})T'+(\dot{B}-\dot{A})\dot{T}\right]f_{TT}+e^{-2A}\Big[\frac{1}{2}\left(A''-B''+\ddot{B}-\ddot{A}\right)\nonumber\\
+\frac{1}{2}e^{2(A+B)}+\left(B'^2-\dot{B}^2+\dot{A}\dot{B}-A'B'\right)+\left(B'-A'\right)e^{A+B}\Big]f_T+\frac{f}{4}=0.
\end{eqnarray}
 First note that in the case of the unperturbed background of Nariai spacetime in which $B=\log(\sqrt{\Lambda}), A=-\log(\sqrt{\Lambda}\cos\tau)$ we find that the scalar torsion  given by (\ref{T2}) is
  \begin{eqnarray}
 T=T_0=-2\Lambda.
 \end{eqnarray}
Therefore, we see that there exists a constant torsion. In the next step, we show that Nariai solution satisfies the ``new" non-diagonal field equations. If we put  $T=T_0+ \delta T(\tau,y),B=\log(\sqrt{\Lambda})+ \delta B(\tau,y),A=-\log(\sqrt{\Lambda}\cos\tau)+ \delta A(\tau,y)$  in the above field equations we observe that the field equations in the non-diagonal formalism is satisfied if the ``generic" function of generalized teleparallel gravity ,``f(T)" satisfies the following constraint equation identically:
\begin{eqnarray}
f(T_0)=0.
 \end{eqnarray}
The last constraint is completely different from the correspondence constraint in the case of diagonal tetrads which has been investigated in details in the previous sections. Here the constraint is just a simple algebraic equation related to the value of $f(T_0)$ and ``not" to it's derivative $f'(T_0)$. For example, here it is easy to show that Nariai solution exists for $f(T)\sim(T-T_0)^{n},\ \ n\in\mathcal{R^{+}}$.\par
Now we derive the perturbation equations of the system around $T=T_0,B=\log(\sqrt{\Lambda}), A=-\log(\sqrt{\Lambda}\cos\tau)$  up to the first order perturbation of fields. The perturbed system reads:
\begin{eqnarray}
\alpha\cos\tau \delta T'-\Lambda\cos\tau\Big[2\sin\tau\delta\dot{B}+\delta B'-\delta A'\Big]-\frac{\delta T}{4}=0\label{per1}\\
\alpha\sec\tau\frac{\delta\dot{T}}{\Lambda}+\Big[\delta\dot{B'}-\tan\tau\delta B'\Big]=0\label{per2}\\
f'(T_0)\Big[\delta\dot{B'}-\tan\tau\delta B'\Big]=0\label{per3}\\
\Lambda\cos^2\tau\Big[\delta\ddot{B}+\sec\tau(\delta B'-\delta A')\Big]+\frac{\delta T}{4}=0\label{per4}\\
-\alpha\cos\tau\Big[\delta T'+\sin\tau\delta\dot{T}\Big]+\frac{\delta T}{4}
+\Lambda\cos^2\tau\Big(\frac{\delta A''-\delta B''+\delta\ddot{B}-\delta\ddot{A}}{2}+\\ \nonumber\sec^2\tau(\delta A+\delta B)+\tan\tau\delta\dot{B}+\sec\tau(\delta B'-\delta A')\Big)=0\label{per5}.
 \end{eqnarray}
In addition, the perturbation of the scalar torsion from (\ref{T2}) reads
\begin{eqnarray}
\delta T=-\frac{2\sec^2\tau}{\Lambda}\Big[\sec^2\tau(3\delta A+2\delta B)+2\sec^2\tau\delta B'+2\tan\tau\delta\dot{B}-2\sec\tau\delta A'\Big]\label{deltaT2}.
 \end{eqnarray}
Because  $f'(T_0)\neq0$, from (\ref{per3}), we obtain
\begin{eqnarray}
\Big[\delta\dot{B'}-\tan\tau\delta B'\Big]=0\label{per33}.
 \end{eqnarray}
By plugging (\ref{per33}) in (\ref{per2}) we find:
\begin{eqnarray}
\delta T=\delta T(y)\label{per22}.
\end{eqnarray}
The integration of (\ref{per33}) gives us :
\begin{eqnarray}
\delta B(\tau,y)=c_1(y)\sec\tau+c_2(\tau)\label{deltaB2}.
\end{eqnarray}
Note that (\ref{deltaB2}) defines the horizon as given by (\ref{relation}) if we have two unknown functions $\{c_1(y),c_2(\tau)\}$. This set of functions must satisfy also the remaining equations (\ref{per1},\ref{per4},\ref{per5}). \par
We add the two equations (\ref{per1},\ref{per5}) to obtain:
\begin{eqnarray}
-\sin\tau\cos\tau\delta\dot{B}+\delta A+\delta B+\frac{\cos^2\tau}{2}(\delta A''-\delta B''+\delta\ddot{B}-\delta\ddot{A})=0\label{perplus}.
\end{eqnarray}
Now we solve (\ref{perplus},\ref{per4}). The full system has the boost symmetry along the killing vector $\partial_x$ so, it is natural that the metric perturbations $\delta A,\delta B$ has a Fourier series for $x\in S^1$ as the following \cite{Bousso:1997wi}:
\begin{eqnarray}
\delta A=\Sigma_{n=1}^{\infty}(\alpha_n(\tau)\cos ny+\beta_n(\tau)\sin ny),\\
\delta B=\sec \tau \Sigma_{n=1}^{\infty}(\alpha'_n\cos ny+\beta'_n\sin ny)+\alpha'_0(\tau).
\end{eqnarray}

By comparison with (\ref{deltaB2}) we find that: 
\begin{eqnarray}
c_1(y)=\Sigma_{n=1}^{\infty}(\alpha'_n\cos ny+\beta'_n\sin ny)=\Sigma_{n=1}^{\infty}\gamma_n\cos(ny-\theta_n),\ \ c_2(\tau)=\alpha'_0(\tau)
\end{eqnarray}
Note that here these two representations are the same just by a simple redefinition $\gamma_n^2=(\alpha'_n)^2+(\beta'_n)^2$ and $\tan\theta_n=\beta'_n/\alpha'_n$. In the next following analysis because we restrict ourselves to the first mode $n=1$ so we will write $\gamma\equiv\gamma_1,\ \ \theta_n=\theta$.\par
To determine uniquely the dynamics of horizon, we need to specify the form of  $\alpha'_0(\tau)$. Through the field equations   (\ref{perplus},\ref{per4}),$\{\alpha'_0(\tau),\alpha_n(\tau),\beta_n(\tau)\}$ coupled stiff non linear. We restrict ourself only to the first mode $n=1$ in which the horizon can be determined analytically. This mode physically is the most alive mode in the system with longest value of wavelength. So, investigation of this first mode is very important and physically viable. The system of differential equations for $\{\alpha'_0(\tau),\alpha_n(\tau),\beta_n(\tau)\}$ is the following (we set $\Lambda=1$):
\begin{eqnarray}
2\, \cos^2\tau \left( {\frac {d}{d\tau}}
\alpha'_0(\tau)  \right) \sin \left( \tau \right) -2\,\alpha'_0(\tau)\cos \tau -{\frac {\alpha'_0(\tau) }{ \left( \cos\tau  \right) ^{3}}}=0\\
-2\alpha_1'(1+\cos^2\tau)+\cos\tau\alpha_1(\tau)(\cos^2\tau-2)+\cos^3\tau\ddot{\alpha_1}=0\\
-2\beta_1'(1+\cos^2\tau)+\cos\tau\beta_1(\tau)(\cos^2\tau-2)+\cos^3\tau\ddot{\beta_1}=0.
\end{eqnarray}
The system is completely integrable in the case of first mode $n=1$. For higher modes we must use superposition of large number of modes. It needs numerical simulation. We will not perform this case in this paper. We would like to know  what will happen in the first mode. The case of a single mode $n>1$ is also interesting. The qualitative behaviour is the same as in the case of the first mode. Therefore, we limit ourselves only to the first mode.
The exact solution for $\alpha'_0(\tau)$ is:
\begin{eqnarray}
\alpha'_0(\tau)=B_0\{\tan\tau\}^{3/2}e^{\frac{1+2\cos^2\tau}{4\cos^4 \tau}}.
\end{eqnarray}
Here $B_0\in\mathcal{R}$ is an arbitrary constants. The first is amplitude of the purely time dependent part of the horizon perturbation and $\gamma$ denotes the amplitude of the inhomogeneous part of the perturbation. Hence, in first mode ``horizon" perturbation is:
\begin{eqnarray}
\delta B(\tau,y)=\gamma\sec \tau \cos(y-\theta)+B_0\{\tan\tau\}^{3/2}e^{\frac{1+2\cos^2\tau}{4\cos^4 \tau}}\label{deltaBfinal}.
\end{eqnarray}
Using (\ref{deltaBfinal}), the location of horizon from (\ref{location}) reads :
\begin{eqnarray}
y_h=-\tau+\theta+\arcsin\{\frac{\cos^2\tau}{\gamma}\frac{d}{d\tau}\alpha'_0(\tau)\},\ \ \{\gamma,\theta\}\in\mathcal{R}.
\end{eqnarray}
This equation gives us the full history of horizon evolutionary scheme in this model under first order small perturbations for long wavelength mode.\par
We plot the time evolution of the location of the horizon $y_h$ in the FIG.1. It shows that the horizon position in the case of the $\gamma>0$ starts from zero and tends to the unit. If we change the sign of the parameter $\gamma$ the behaviour is symmetrically reversed.
\begin{figure}
\begin{center}
\includegraphics[angle=0, width=0.5\textwidth]{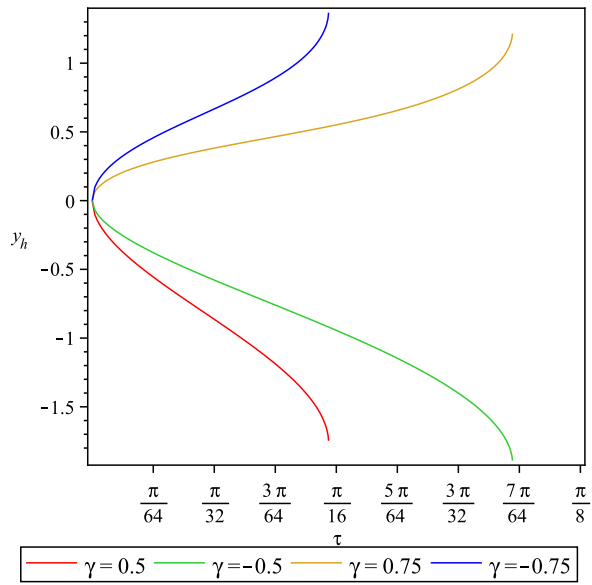}
\end{center}
\caption{\label{Fig2a} Location of horizon $y_h$ for $n=1$ and different choice of $\gamma,B_0=1/2$.}
\end{figure}

Thus, we have:
\begin{eqnarray}
\delta B(\tau)=\gamma\sec \tau \cos(\tau-\arcsin\{\frac{\cos^2\tau}{\gamma}\frac{d}{d\tau}\alpha'_0(\tau)\})+\alpha'_0(\tau)\label{deltaBtau}.
\end{eqnarray}
We have to see that (\ref{deltaBtau}) is increasing(anti evaporating) or decreasing (evaporation); specially we must show that it is divergence when $\tau\rightarrow\frac{\pi}{2}$.

\begin{figure}
\begin{center}
\includegraphics[angle=0, width=0.5\textwidth]{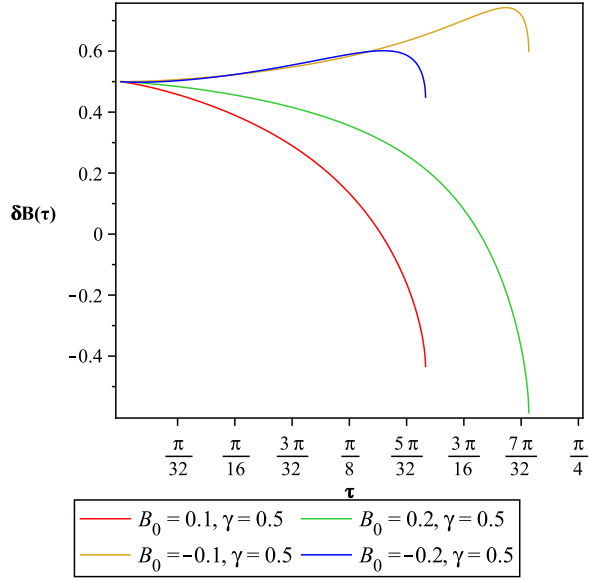}
\end{center}
\caption{\label{Fig2a} Evolution of (\ref{deltaBtau}) for $n=1$ and different choice of $\gamma,B_0$.}
\end{figure}
The quantitative behaviour of the horizon perturbation has been plotted in the FIG.1. As we observe, when the set of the parameters of the model $B_0,\gamma$ changes, we can have an increasing or decreasing behaviours. The final stage of the black hole perturbation for grey-blue graphs corresponds to the evaporation. But the behaviour is different when we choose another set of data as red-green ones. In this case the anti-evaporation is the final destination of the black hole. So, in spite of the diagonal analysis, in the non-diagonal formalism we can have both evaporation and anti-evaporation . The quantitative behaviour is a generic one. \par
Because in comparison of GR and other modified gravities \cite{Lorenzo} both phenomena must happen in a same background, depending on the parameters and initial perturbation data, now, in a non-diagonal case we found the same phenomena. So, from the dynamical point of view, also we prefer the non-diagonal tetrads because the physical phenomena is reasonable and physically acceptable. We must have both behaviours for a typical model of black hole in this torsion based model. Absence of evaporation means we must change our tetrads formalism from diagonal to non-diagonal, as we did here. So, these calculations complete our arguments about the evaporation in $f(T)$.
\par
Moreover, we mention that the choice of a set of non-diagonal tetrads is quick consistent with the dependence of the referential under consideration \cite{manusug1,manusug2} and a freedom in the choice of the algebraic function $f(T)$, where $f_{TT}\neq 0$ for a non-constant torsion scalar. It is also important to note that for a non-constant torsion scalar, Eqs.~(\ref{david21})-(\ref{david22}) lead to the constraint $f_{TT}=0$, while  from Eqs.~(\ref{david49})-(\ref{david50}) one has a freedom on the choice of the algebraic function $f(T)$ where in general  $f_{TT}$ is different from zero. Further, we mention that the tetrad chosen in this paper is a good. This kind of work in order to find good tetrads (the non-diagonal ones) has been performed in \cite{bohemer}, but within Friedmann-Robertson-walker universe, showing the importance of using such tetrads.

 \section{Conclusion}
In brief, we investigated the existence of evaporation in the $f(T)$ gravity using two different kinds of tetrad, the usual diagonal and also the non-diagonal one. By starting from a diagonal tetrads, in spite of the Einstein gravity and another modified gravities based on curvature, when you are working in a Weitzenbock spacetime, with torsion, we explicitly showed that (anti)evaporation never happen. The absence of evaporation is an universal property of $f(T)$ gravity and it is independent from the generic form of $f(T)$ models. But in $f(R)$ gravity you can have both evaporation/anti-evaporation for viable models. This result implies that a typical black hole in any generic form of generalized teleparallel gravity is frozen in it's initial state if we use the diagonal tetrads.  But the story is completely different if we use the non-diagonal tetrads. With the non-diagonal tetrads, we have both the (anti)evaporation dependence on the parameters of the  $f(T)$ model and also the initial horizon perturbation phase. Moreover, our results show the importance of using non-diagonal tetrads where we note that Eqs.~(\ref{david21})-(\ref{david22}) yield a null second derivative of the algebraic function $f(T)$, i.e., $f_{TT}=0$, while  from Eqs.~(\ref{david49})-(\ref{david50}) one has a freedom on the choice of the algebraic function $f(T)$ where in general  $f_{TT}$ is different from zero. All this interesting result comes from the fact that in this  paper we have made a good choice of the non-diagonal tetrad. We conclude that it is important to look for suitable choice  of non-diagonal tetrads as performed in \cite{bohemer} in order to scape from a constrained expression for the algebraic function $f(T)$.

\vspace{0.5cm}
{\bf Acknowledgement:}
The authors thank Prof.~S.~D.~Odintsov for having suggested the topic of
research and also for useful comments.  M. E. Rodrigues thanks the UFES and UFPA for the hospitality during the development of this work and  thanks CNPq for partial financial support.  M. J. S. Houndjo  thanks CNPq/FAPES for financial support.


\begin{thebibliography}{90}

\bibitem{davis}
P.C.W. Davies, Rept.Prog.Phys. 41 (1978) 1313-1355.

\bibitem{Bousso:1997wi} 
  R.~Bousso, S.~W.~Hawking and ,
  Phys.\ Rev.\ D {\bf 57}, 2436 (1998)
  [hep-th/9709224].



\bibitem{Nojiri:2010wj}
  S.~'i.~Nojiri, S.~D.~Odintsov and ,
  Phys.\ Rept.\  {\bf 505} (2011) 59
  [arXiv:1011.0544 [gr-qc]].
\bibitem{Nojiri:2008ku}
  S.~'i.~Nojiri, S.~D.~Odintsov and ,
  AIP Conf.\ Proc.\  {\bf 1115} (2009) 212
  [arXiv:0810.1557 [hep-th]].
\bibitem{Nojiri:2008nt}
  S.~'i.~Nojiri, S.~D.~Odintsov and ,
  TSPU Bulletin N {\bf 8(110)} (2011) 7
  [arXiv:0807.0685 [hep-th]].
\bibitem{Nojiri:2006ri}
  S.~'i.~Nojiri, S.~D.~Odintsov and ,
  eConf C {\bf 0602061} (2006) 06
   [Int.\ J.\ Geom.\ Meth.\ Mod.\ Phys.\  {\bf 4} (2007) 115]
  [hep-th/0601213].








\bibitem{Nojiri:2013su} 
  S.~'i.~Nojiri, S.~D.~Odintsov and ,
  arXiv:1301.2775 [hep-th].
  \bibitem{Nariai}
H. Nariai. 1950. Sci.Rept.Tohoku Univ.(Ser.A),34,160; H. Nariai. 1951. Sci.Rept.Tohoku Univ.(Ser.A),35,62.

\bibitem{fT}
 G.~R.~Bengochea, R.~Ferraro and ,
  Phys.\ Rev.\ D {\bf 79}, 124019 (2009)
  [arXiv:0812.1205 [astro-ph]].
  
\bibitem{Linder}
   E.~V.~Linder,
  Phys.\ Rev.\ D {\bf 81}, 127301 (2010)
  [Erratum-ibid.\ D {\bf 82}, 109902 (2010)]
  [arXiv:1005.3039 [astro-ph.CO]].
  
\bibitem{Jamil:2012ti}
  M.~Jamil, D.~Momeni and R.~Myrzakulov,
  Eur.\ Phys.\ J.\ C {\bf 72} (2012) 2267
  [arXiv:1212.6017 [gr-qc]].
\bibitem{Myrzakulov:2012sp}
  R.~Myrzakulov,
  Entropy {\bf 14} (2012) 1627
  [arXiv:1212.2155 [gr-qc]].
\bibitem{Setare:2012ry}
  M.~R.~Setare and N.~Mohammadipour,
  JCAP {\bf 1211} (2012) 030
  [arXiv:1211.1375 [gr-qc]].
\bibitem{Jamil:2012nm}
  M.~Jamil, D.~Momeni, R.~Myrzakulov and P.~Rudra,
  J.\ Phys.\ Soc.\ Jap.\  {\bf 81} (2012) 114004
  [arXiv:1211.0018 [physics.gen-ph]].
\bibitem{Sadjadi:2012xa}
  H.~M.~Sadjadi,
  Phys.\ Lett.\ B {\bf 718} (2012) 270
  [arXiv:1210.0937 [gr-qc]].
\bibitem{Rodrigues:2012qua}
  M.~E.~Rodrigues, M.~J.~S.~Houndjo, D.~Saez-Gomez and F.~Rahaman,
  Phys.\ Rev.\ D {\bf 86} (2012) 104059
  [arXiv:1209.4859 [gr-qc]].
\bibitem{Jamil:2012ju}
  M.~Jamil, D.~Momeni and R.~Myrzakulov,
  Eur.\ Phys.\ J.\ C {\bf 72} (2012) 2122
  [arXiv:1209.1298 [gr-qc]].
\bibitem{Myrzakulov:2012qp}
  R.~Myrzakulov,
  Eur.\ Phys.\ J.\ C {\bf 72} (2012) 2203
  [arXiv:1207.1039 [gr-qc]].
\bibitem{Houndjo:2012sz}
  M.~J.~S.~Houndjo, D.~Momeni and R.~Myrzakulov,
  Int.\ J.\ Mod.\ Phys.\ D {\bf 21} (2012) 1250093
  [arXiv:1206.3938 [physics.gen-ph]].
\bibitem{Rodrigues:2012wt}
  M.~E.~Rodrigues, M.~H.~Daouda and M.~J.~S.~Houndjo,
  arXiv:1205.0565 [gr-qc].
\bibitem{Setare:2012vs}
  M.~R.~Setare and M.~J.~S.~Houndjo,
  arXiv:1203.1315 [gr-qc].
\bibitem{Bamba:2012rv}
  K.~Bamba, M.~Jamil, D.~Momeni and R.~Myrzakulov,
  arXiv:1202.6114 [physics.gen-ph].
\bibitem{Bamba:2012vg}
  K.~Bamba, R.~Myrzakulov, S.~'i.~Nojiri and S.~D.~Odintsov,
  Phys.\ Rev.\ D {\bf 85} (2012) 104036
  [arXiv:1202.4057 [gr-qc]].


\bibitem{Jamil:2012ti}
  M.~Jamil, D.~Momeni and R.~Myrzakulov,
  Eur.\ Phys.\ J.\ C {\bf 72} (2012) 2267
  [arXiv:1212.6017 [gr-qc]].
\bibitem{Jamil:2012ck}
  M.~Jamil, D.~Momeni and R.~Myrzakulov,
  Gen.\ Rel.\ Grav.\  {\bf 45} (2013) 263
  [arXiv:1211.3740 [physics.gen-ph]].
\bibitem{Jamil:2012ju}
  M.~Jamil, D.~Momeni and R.~Myrzakulov,
  Eur.\ Phys.\ J.\ C {\bf 72} (2012) 2122
  [arXiv:1209.1298 [gr-qc]].
\bibitem{Jamil:2012vb}
  M.~Jamil, D.~Momeni and R.~Myrzakulov,
  Eur.\ Phys.\ J.\ C {\bf 72} (2012) 2075
  [arXiv:1208.0025 [gr-qc]].
\bibitem{Jamil:2012yz}
  M.~Jamil, K.~Yesmakhanova, D.~Momeni and R.~Myrzakulov,
  Central Eur.\ J.\ Phys.\  {\bf 10} (2012) 1065
  [arXiv:1207.2735 [gr-qc]].
\bibitem{Houndjo:2012sz}
  M.~J.~S.~Houndjo, D.~Momeni and R.~Myrzakulov,
  Int.\ J.\ Mod.\ Phys.\ D {\bf 21} (2012) 1250093
  [arXiv:1206.3938 [physics.gen-ph]].
\bibitem{Jamil:2012nm}
  M.~Jamil, D.~Momeni and R.~Myrzakulov,
  Eur.\ Phys.\ J.\ C {\bf 72} (2012) 1959
  [arXiv:1202.4926 [physics.gen-ph]].




  
\bibitem{Daouda:2012nj}
  M.~H.~Daouda, M.~E.~Rodrigues and M.~J.~S.~Houndjo,
  Phys.\ Lett.\ B {\bf 715} (2012) 241
  [arXiv:1202.1147 [gr-qc]].
\bibitem{Jamil:2012zm} 
  M.~Jamil, S.~Ali, D.~Momeni, R.~Myrzakulov and ,
  Eur.\ Phys.\ J.\ C {\bf 72}, 1998 (2012)
  [arXiv:1201.0895 [physics.gen-ph]].
  \bibitem{Jamil:2011mc} 
  M.~Jamil, D.~Momeni, N.~S.~Serikbayev, R.~Myrzakulov and ,
  Astrophys.\ Space Sci.\  {\bf 339}, 37 (2012)
  [arXiv:1112.4472 [physics.gen-ph]].
  \bibitem{Jamil:2011iu} 
  M.~Jamil, D.~Momeni, M.~A.~Rashid and ,
  Eur.\ Phys.\ J.\ C {\bf 71}, 1711 (2011)
  [arXiv:1107.1558 [physics.gen-ph]].
  \bibitem{Hendi:2012nj} 
  S.~H.~Hendi, D.~Momeni and ,
  Eur.\ Phys.\ J.\ C {\bf 71}, 1823 (2011)
  [arXiv:1201.0061 [gr-qc]].
  

\bibitem{Daouda:2011yf}
  M.~Hamani Daouda, M.~E.~Rodrigues and M.~J.~S.~Houndjo,
  Eur.\ Phys.\ J.\ C {\bf 72} (2012) 1893
  [arXiv:1111.6575 [gr-qc]].
\bibitem{Daouda:2011rt}
  M.~Hamani Daouda, M.~E.~Rodrigues and M.~J.~S.~Houndjo,
  Eur.\ Phys.\ J.\ C {\bf 72} (2012) 1890
  [arXiv:1109.0528 [physics.gen-ph]].
\bibitem{HamaniDaouda:2011iy}
  M.~Hamani Daouda, M.~E.~Rodrigues and M.~J.~S.~Houndjo,
  Eur.\ Phys.\ J.\ C {\bf 71} (2011) 1817
  [arXiv:1108.2920 [astro-ph.CO]].
\bibitem{Myrzakulov:2010tc}
  R.~Myrzakulov,
  Gen.\ Rel.\ Grav.\  {\bf 44} (2012) 3059
  [arXiv:1008.4486 [physics.gen-ph]].
\bibitem{Yerzhanov:2010vu}
  K.~K.~Yerzhanov, S.~.R.~Myrzakul, I.~I.~Kulnazarov and R.~Myrzakulov,
  arXiv:1006.3879 [gr-qc].
\bibitem{Myrzakulov:2010vz}
  R.~Myrzakulov,
  Eur.\ Phys.\ J.\ C {\bf 71} (2011) 1752
  [arXiv:1006.1120 [gr-qc]].
\bibitem{Rodrigues:2013uua} 
  M.~E.~Rodrigues, M.~J.~S.~Houndjo, D.~Momeni, R.~Myrzakulov and ,
  arXiv:1302.4372 [physics.gen-ph].
  \bibitem{Emparan}
   R.~Emparan, H.~S.~Reall and ,
  Living Rev.\ Rel.\  {\bf 11}, 6 (2008)
  [arXiv:0801.3471 [hep-th]].
  \bibitem{Lorenzo}
  L.Sebastiani et.al, (In preparation).

\bibitem{manusug1} Baojiu~Li, T.~P.~Sotiriou and J.~D.~Barrow, Phys.~Rev.~D {\bf 83}, 064035~(2011); Phys.~Rev.~D {\bf 83}, 104030 (2011).

\bibitem{manusug2} C.~Deliduman and B.~Yapiskan, ``Absence of Relativistic Stars in f(T) Gravity",  arXiv:1103.2225v3~[gr-qc]. 

\bibitem{bohemer} N. Tamanini and C. G. Boehmer, Phys. Rev. D {\bf 86}, 044009 (2012),  	arXiv:1204.4593 [gr-qc].



\end{thebibliography}
\end{document}